\documentstyle[12pt]{article}
\newcommand{\beq}{\begin{equation}}
\newcommand{\eeq}{\end{equation}}
\newcommand{\bea}{\begin{eqnarray}}
\newcommand{\eea}{\end{eqnarray}}

\newcommand{\lsim}{\stackrel{\scriptstyle <}{\phantom{}_{\sim}}}

\newcommand{\Ox}{^{24}\mbox{O}}
\newcommand{\Oxx}{^{26}\mbox{O}}

\begin{document}

\centerline{\large\bf On the nucleon instability of heavy oxygen isotopes}

\vskip 0.7 cm

\centerline{Yu.~S.~Lutostansky$^1$ and M.~V.~Zverev$^2$}

\vskip 0.5 cm

\centerline{\small $^1$ Moscow Engineering Physics Institute,
115409, Moscow, Russia }
\centerline{\small $^2$ Kurchatov Institute, 123182, Moscow, Russia }

\vskip 1 cm

\begin{abstract}
The instability of the $\Oxx$ nucleus with respect to decay through
the two-neutron channel is investigated. It is shown that this
isotope, unobserved in the fragmentation experiments, can exist
as a narrow resonance in the system $\Ox+2n$. A role of deformation in
formation of the neutron-drip line in the region $N\sim 20$ is
discussed.

\end{abstract}

\vskip 1 cm

The structure of nucelei near the neutron-drip line possesses
interesting features. These are: 1) two-neutron instability
of nuclei which are stable with respect to one-neutron emission,
2) arising a new region of deformation when approaching the
neutron-drip line, 3) change of the concept of nuclear shells at
the boundary of neutron stability, 4) formation of the neutron
halo, 5) clasterization in very neutron-rich nuclei. In this
article, we pay attention to nucleon stability of oxygen nuclei
near the neutron-drip line. This region, close to double magic
area, has been attracting an attention of nuclear
experimentalists an theorists for last decade.

The instability of the $\Oxx$ nucleus has been observed for
the first time at the LISE spectrometer at GANIL
in the fragmentation of the $^{48}$Ca beam with
an energy of 44 Mev per nucleon on a Ta target \cite{ganil}.
The work \cite{michigan} confirmed this result. No events
associated with the isotope $\Oxx$ were also observed in
the recent analysis of oxygen fragments by the separator
RIPS in the projectile fragmentation experiments
with using 94.1 MeV per nucleon $^{40}$Ar beam at the
RIKEN \cite{riken}. Theoretical study
of properties of $\Oxx$ was done in \cite{yaf} within the
framework of the self-consistent theory of finite Fermi systems
\cite{migdal,zs}. Estimation of one-neutron separation energy
yielded $S_n\simeq 0.7\,$MeV, i.e. this nucleus was found to be
stable to one-neutron emission. This clears the way for the pure
two-neutron instability of the $\Oxx$ nucleus. Before going over
to investigation of this subject we concern briefly neighbouring
even-even oxygen isotopes. Calculations made in \cite{yaf} for the
nucleus $\Ox$ observed in all the experiments yielded
$S_n=3.6\,$MeV and $S_{2n}=5.9\,$MeV. At the same time those
calculations showed 1$n$- and 2$n$-instability of the nucleus $^{28}$O
unobserved in \cite{ganil,michigan,riken}.

The two-neutron instability corresponding to positive
one-neutron separation energy $S_n$ and negative
two-neutron separation energy $S_{2n}$ is possible owing
to the positive energy of neutron pairing \cite{gold}.
This energy is usually characterized by the even-odd difference
$\Delta_n$ of the binding energies $E_b$ of three neighboring
isotopes, which is given by
\beq \label{delta}
\Delta_n =\frac{1}{2}\Bigl[E_b(N{+}2,Z)-2E_b(N{+}1,Z)
+E_b(N,Z)\Bigr]
\eeq
and is positive for even-even nucleus $(N,Z)$. The obvious
relation $S_{2n}-2S_n=-2\Delta_n$ follows from eq.~(\ref{delta}).
This relation indicates that two-neutron instability is possible
provided that the condition $0{<}S_n{<}\Delta_n$ is satisfied.
Bearing in mind that in this region of nuclei
$\Delta_n\sim 2\div 3\,$MeV one can see that this condition
holds for the $\Oxx$ nucleus.

To investigate the stability of the $\Oxx$ nucleus, we consider
the $\Ox+2n$ system. The energies $E_q$ and widths $\Gamma_q$
of quasistationary states of this system are determined by the
poles $E_q-i\Gamma/2$ of the Green's function of two neutrons in
the mean field of the $\Ox$ nucleus \cite{migdal}. The equation
for the two-neutron Green's function can be reduced
to the equation
for the two-neutron wave function $\Psi(1,2)$:
\beq \label{equa}
\Bigl[H_0(1)+H_0(2)+V(1,2)\Bigr]\Psi(1,2) = E\Psi(1,2),
\eeq
where the arguments 1 and 2 stand for the sets of neutron
spatial and spin coordinates, $H_0$ is the single-particle
Hamiltonian, and $V$ is the effective interaction in the
two-particle channel. Recent microscopic calculation of this
interaction for semi-infinite nuclear matter \cite{semi}
showed its surface nature. The phenomenological
analysis of even-odd effects in tin isotopic chain
\cite{fayans} showed that density dependence of the effective
pairing interaction is quite sophisticated. However the
parameters controlling the shape and the strength of this
interaction are fitted for rather heavy nuclei not very
far from the valley of $\beta$-stability and can hardly be
directly used in the problem under discussion. That is why
we did not claim to calculate the wave function $\Psi(1,2)$
and the energy $E$. The aim is to estimate the width of
the two-neutron state by the order of value.

To solve eq.~(\ref{equa}), we expand the function $\Psi(1,2)$
in the basis formed by the products of two eigenfunctions of the
Hamiltonian $H_0$ with the total angular momentum $I{=}0$
and its projection $M{=}0$. This expansion has the form
\beq \label{expa}
\Psi(1,2) = \int\limits_0^{\infty}\!\int\limits_0^{\infty}
d\varepsilon_1 d\varepsilon_2 C^{\nu}(\varepsilon_1,\varepsilon_2)
\Bigl\{\psi^{\nu}_{\varepsilon_1}(1) \cdot
\psi^{\nu}_{\varepsilon_2}(2)\Bigr\}^{00},
\eeq
where $\nu$ is the set of angular quantum numbers.

For low energies, the $\varepsilon$-dependence of the functions
$\psi_{\varepsilon}(1)$ can be separated from the coordinate
and spin dependences as follows \cite{gali,perelom}:
\beq \label{sepa}
\psi_{\varepsilon}(1) = \sqrt{\Delta(\varepsilon)}\psi_0(1).
\eeq

According to the theoretical scheme of neutron single-partcle
levels of the $\Ox$ nucleus, the $2s_{1/2}$ state is the
shallowest bound state, and the $1d_{3/2}$ state is the lowest
quasistationary state. At low energies, the factor
\beq \label{factd}
\Delta^d(\varepsilon) = \frac{1}{2\pi}\,
\frac{\gamma}{(\varepsilon - \varepsilon_0^d)^2+\gamma^2/4}
\eeq
is resonantly enhanced for the $d_{3/2}$ continuum states
because of proximity to the $1d_{3/2}$ quasistationary state
with energy $\varepsilon^d=\varepsilon^d_0{-}i\gamma/2$ in the
$\Ox$ nucleus \cite{gali}. Owing to existence of the weakly
bound $2s_{1/2}$ state, the wave functions of the $s_{1/2}$
continuum states are enhanced at low energies by the factor
\cite{perelom}
\beq \label{facts}
\Delta^s(\varepsilon) \simeq \frac{\sqrt{2m}\beta}
{\pi\sqrt{\varepsilon}(\varepsilon{+}\beta)},
\eeq
where $\beta{=}2/mR^2$ and $R$ is of the order of the $\Ox$
radius. For this reasons, we restrict the basis in the
expansion (\ref{expa})
of the wave function $\Psi(1,2)$ to the $s_{1/2}$ and
$d_{3/2}$ states of single-particle continuum.

To derive the required relations, we follow the method used in
\cite{gali} to obtain the relations of the theory of
two-proton radioactivity. Substituting expansion (\ref{expa})
with the functions $\psi_{\varepsilon}(1)$ factorized in form
(\ref{sepa}) into
eq.~(\ref{equa}), we find that the energy of the quasistationary
state satisfies the algebraic equation
\beq \label{alge}
g_{dd}\,M^d(E)+g_{ss}\,M^s(E)+(g_{ds}^2{-}g_{dd}\,g_{ss})\,
M^d(E)\,M^s(E)=1,
\eeq
where
\vskip -0.6 cm
\beq \label{mmm}
M^{\lambda}(E)=\int\limits_0^{\infty}\!\int\limits_0^{\infty}
d\varepsilon_1 d\varepsilon_2\;
\frac{\Delta^{\lambda}(\varepsilon_1)\Delta^{\lambda}(\varepsilon_2)}
{\varepsilon_1{+}\varepsilon_2{-}E},
\eeq
\beq \label{ggg}
g_{\lambda\nu} = -\int d1\,d2\,V(1,2)\,
\Bigl\{\psi^{\lambda}_0(1) \cdot \psi^{\lambda}_0(2)\Bigr\}^{00}
\Bigl\{\psi^{\nu}_0(1) \cdot \psi^{\nu}_0(2)\Bigr\}^{00},
\eeq
and $\lambda,\nu=s,d$. Substituting expresiions (\ref{factd})
and (\ref{facts}) for the quantity $\Delta^{\lambda}(\varepsilon)$
into eq.~(\ref{mmm}) and calculating the integrals by rotating the
contour to the negative imaginary axis, we obtain
\beq \label{ms}
M^s(E) = 2m\left(-{1\over\pi}\ln{E\over 4\beta}-{1\over 2} +i\right),
\eeq
\beq \label{md}
M^d(E)= \frac{1}{2\varepsilon_0^d{-}E}+
\frac{2\gamma^2}{\pi(2\varepsilon_0^d{-}E)^2}
\left\{ \frac{E}{\varepsilon_0^d{-}E} +
\frac{1}{2\varepsilon_0^d{-}E}
\ln\left\vert\frac{\varepsilon_0^d}{\varepsilon_0^d{-}E}
\right\vert\right\}.
\eeq

Eq.~(\ref{alge}) with $M^s(E)$ and $M^d(E)$ specified by
eqs.~(\ref{ms}) and (\ref{md}) is an algebraic equation with
complex coefficients. Numerical analysis of eq.~(\ref{alge})
shows that it has a complex root $E=E_0-i\Gamma/2$ in a wide
region of input parameters. While the position of this root
depends on the values of the parameters, it is located around
$E_0\sim 0.1\,$MeV, $\Gamma\sim 10^{-3}\,$MeV. We reserve a
detailed study of this dependence for the future publication.
The width $\Gamma\sim 10^{-3}\,$MeV of this quasistationary state
corresponds to a lifetime $\tau\sim\hbar/\Gamma\sim 10^{-18}\,$s.
Since $\tau\gg\tau_{nucl}\sim 10^{-22}\,$s the emitted neutron pair
should be a weakly bound di-neutron state. It can be observed
in correlation experiments. The similar situation
was discussed earlier in connection with an analyzis of the
$\beta$-delayed multi-neutron emission \cite{lp,lsp}. However,
in that case the cascade ($n{+}n$) neutron emission strongly dominates
and the di-neutron ($^2n$) channel is suppressed (e.g. for $^{35}$Na,
$P_{^2n}/P_{n{+}n}< 0.19$ \cite{lsp}) due to a decay of the excited
state. In the case of $\Oxx$, the neutron pair is emitted from the
ground state, so that the situation is ``more clear''.

The spherical basis was used in calculations for neutron-rich
oxygen isotopes. However, account of deformation (even small,
with $\beta_2<0.2$) could strongly change the picture. Indeed,
splitting of the single-particle $d_{3/2}$ neutron state results in
destroying the shell $N=20$ for nuclei near the neutron-drip
line \cite{nfs5}.
Such a sensitivity to small deformations should essentially
complicate description of neutron-rich nuclei. However, the oxygen
isotopes evidently should not be deformed. Otherwise, being deformed
with $\beta_2>0.1$, the isotope $\Oxx$ could obtain additional
stability due to increase of the binding energy by $\sim$ 1 MeV
and should be observed experimentally, but this is not the case.

This is the deformation that seems to give an explanation of
existence of the isotope $^{31}$F observed in the experiment
\cite{riken}. The calculations \cite{yaf} of one- and
two-neutron separation energies for this nucleus based on the
self-consistent finite Fermi system theory in the sphericl
geometry yielded nucleon instability of $^{31}$F. At the same
time the scenario of sharp shift of the neutron-drip line owing
to onset of a deformation was suggested several years ago for
explanation of the nucleon stability of heavy sodium isotopes
\cite{detraz}. Nuclei with less than half occupation of the
level $f_{7/2}$ being slightly unbound in spherical calculation
($S_n\lsim 0$) can become stable due to lowering the energy of states
with asymptotic quantum numbers $\frac{1}{2}^-[330]$ and
$\frac{3}{2}^-[321]$ at deformation \cite{izv}. The analogous
scenario seems to take place for fluorine isotopes: the neutron level
$f_{7/2}$ just starts to be occupied in the nucleus $^{31}$F
with $N=22$. Following this scenario one could expect nucleon
stability of the isotope $^{33}$F as well. The problem of onset of a
deformation for nuclei near the neutron-drip line will be discussed
in a separate article, in particular, in connection with an opportunity
of clasterization in weakly bound neutron-rich systems.

\vskip 0.3 cm
Authors are grateful to B.~V.~Danilin, D.~Guillemaud-Mueller
and M.~V.~Zhukov for valuable discusiions.


\end{document}